\documentclass[twocolumn,pra,showpacs,amsmath]{revtex4-1}
\def\qed{\leavevmode\unskip\penalty9999 \hbox{}\nobreak\hfill
     \quad\hbox{\leavevmode  \hbox to.77778em{%
               \hfil\vrule   \vbox to.675em%
               {\hrule width.6em\vfil\hrule}\vrule\hfil}}
     \par\vskip3pt}
\usepackage{amsmath}
\newtheorem{theorem}{Theorem}

\newtheorem{lemma}{Lemma}
\begin{document}

\title{Maximally entangled state and fully entangled fraction}
\author{Ming-Jing Zhao}
\affiliation{Department of Mathematics, School of Science,
Beijing Information Science and Technology University, 100192, Beijing, China.}

\begin{abstract}
We study maximally entangled states and fully entangled fraction in general $d^\prime\otimes d$ ($d^\prime\geq d$) systems.
Necessary and sufficient conditions for maximally entangled pure and mixed states are presented.
As a natural generalization of the usual fully entangled fraction for $d\otimes d$ systems,
we define the maximal overlap between a given quantum state and the maximally entangled states
as the fully entangled fraction in $d^\prime\otimes d$ systems. The properties of this fully entangled fraction and
its relations to quantum teleportation have been analyzed. The witness for detecting maximally entangled states
and quantum states that are useful for quantum teleportation is provided.
\end{abstract}

\maketitle

\section{Introduction}

Entanglement is a vital resource in quantum information processing. In particular, the maximally entangled states are
believed to be the ideal resource in many quantum information processing tasks \cite{M. A. Nielsen, R. Horodecki}.
Much efforts have been devoted to increase the degree of entanglement in quantum states. Actually it has been
proved that all maximally entangled states are pure in bipartite $d\otimes d$ systems \cite{D. Cavalcanti}. However, in Ref. \cite{Z. G. Li}
a class of mixed states which are also maximally entangled has been introduced. This special class of mixed states is shown to be
the ideal resource for quantum teleportation.

One quantity tightly related to maximally entangled states is the
fully entangled fraction, which plays important role in teleportation
\cite{C. H. Bennett1993}.
For instance the
fidelity of optimal teleportation is given by fully entangled fraction \cite{C. H. Bennett1996,
M. Horodecki1999,S. Albeverio}.
Additionally, the fully entangled fraction in two-qubit system acts
as an index characterizing the nonlocal correlation \cite{Z. W. Zhou} and plays a significant
role in deriving two bounds on the damping rates of the dissipative
channel \cite{SKO}.

Since fully entangled fraction has clear experimental and theoretical significance, an
analytic formula for fully entangled fraction is of great importance. In
Ref. \cite{J. Grondalski} an elegant formula for fully entangled fraction in two-qubit system is
derived analytically by using the method of Lagrange multiplier. For
high dimensional quantum states the analytical computation of fully entangled fraction
is formidably difficult, and less results have been known. In
Ref. \cite{upperbound} the upper bound of the fully entangled fraction has been estimated.
In Ref. \cite{M. J. Zhao2010} some analytical results have been derived for some special states.
The monogamy relations in terms of fully entangled fraction have been proven for multiqubit pure states,
but it is not true for general mixed states \cite{S. Lee}.

Since the fully entangled fraction of $d\otimes d$ quantum states is the maximal overlap with the maximally entangled pure states,
it is only well defined for bipartite systems with the subsystems of the same dimensions.
One question is whether there is a similar quantity as fully entangled fraction in general bipartite systems,
and what the theoretical or experimental meanings of such quantities are.
Fortunately, the existence of maximally entangled mixed states in general bipartite systems provides us insight to answer this question.

In this paper, we mainly investigate maximally entangled states and fully entangled fraction in general $d^\prime\otimes d$ $(d^\prime\geq d)$ systems.
Without loss of generality, we assume $d^\prime=Kd+r$, $0\leq r< d$, $K\geq 1$.
First, we give necessary and sufficient conditions of maximally entangled pure and mixed states.
Based on these results, we define the maximal overlap between a given quantum state and the maximally entangled states
as the fully entangled fraction in $d^\prime \otimes d$ systems, which is a natural generalization of the usual fully entangled fraction.
The properties of this fully entangled fraction and its relations to quantum teleportation is analyzed,
to show that the fully entangled fraction is meaningful for general bipartite systems.
Finally the witness for detecting maximally entangled states and the quantum states that are useful for quantum teleportation is provided.

\section{Maximally entangled states}

In Ref. \cite{Z. G. Li}, it has been shown that the maximally entangled state can be either pure or mixed.
In $d^\prime \otimes d$ system, if a pure state is maximally entangled, then its Schmidt coefficients are all
equal to $\frac{1}{\sqrt{d}}$. If a mixed state is maximally entangled, then it is a convex combination of maximally entangled pure states that are pairwise orthogonal with each other.
\begin{lemma}\label{lemma max}
A $d^\prime \otimes d$ bipartite mixed state $\rho$ is maximally entangled if and only if
\begin{eqnarray}
\rho=\sum_{m=1}^K p_m |\psi_m\rangle \langle \psi_m|,\ \ \sum_{m=1}^K p_m=1,
\end{eqnarray}
where
\begin{eqnarray}
|\psi_m\rangle=\frac{1}{\sqrt{d}}\sum_{i=0}^{d-1} |f_{im}\rangle\otimes |e_i\rangle,
\end{eqnarray}
with $\{|f_{im}\rangle\}_{im}$ and $\{|e_i\rangle\}_i$ satisfying $\langle f_{i^\prime m^\prime}|f_{im}\rangle=\delta_{ii^\prime}\delta_{mm^\prime}$ and $\langle e_{i^\prime}|e_i\rangle=\delta_{ii^\prime}$ .
\end{lemma}

In the following, we will show two necessary and sufficient conditions for maximally entangled states.
We denote $d$ dimensional vector $|i\rangle$ as the vector having only one nonzero entry 1 in the $(i-1)$-th position and $d^\prime$ dimensional vector $|i+(m-1)d\rangle$ as the vector having only one nonzero entry 1 in the $i\times (m-1)d-1$-th position, $i=0,\cdots,d-1$, $m=1,\cdots,K$. In this way, $\{|i\rangle\}$ and $\{|i+(m-1)d\rangle\}$ are orthonormal bases of the $d$ dimensional Hilbert space and $Kd$ dimensional Hilbert space respectively. Define
\begin{eqnarray}\label{max ent pure}
|\chi_m\rangle\equiv\frac{1}{\sqrt{d}}\sum_{i=0}^{d-1} |i+(m-1)d\rangle\otimes |i\rangle.
\end{eqnarray}
$\{|\chi_m\rangle\}$ are maximally entangled pure states that are pairwise orthogonal.
Employing these maximally entangled states $\{|\chi_m\rangle\}$, we introduce the first necessary and sufficient condition for maximally entangled state.

\begin{theorem}\label{th all max ent carnonical form}
$\rho$ is maximally entangled if and only if there exists unitary operator $U$ acting on the first subsystem such that
\begin{eqnarray}\label{all max ent carnonical form }
(U\otimes I)\,\rho\, (U^\dagger \otimes I) =\sum_{m=1}^K p_m |\chi_m\rangle \langle \chi_m|,
\end{eqnarray}
where $p_m$ is the eigenvalue of $\rho$, $p_m\geq 0$, $\sum_{m=1}^K p_m=1$.
\end{theorem}

Proof. First, if a quantum state $\rho$ satisfies Eq. (\ref{all max ent carnonical form }), then it is obvious that both $\sum_{m=1}^K p_m |\chi_m\rangle \langle \chi_m|$ and $\rho$ are maximally entangled.

Second, suppose $\rho$ is maximally entangled, i.e. $\rho=\sum_{m=1}^K p_m |\psi_m\rangle \langle \psi_m|$, $\sum_{m=1}^K p_m=1$,
$|\psi_m\rangle=\frac{1}{\sqrt{d}}\sum_{i=0}^{d-1}  |f_{im}\rangle\otimes |e_i\rangle$
with $\langle f_{i^\prime m^\prime}|f_{im}\rangle=\delta_{ii^\prime}\delta_{mm^\prime}$ and $\langle e_{i^\prime}|e_i\rangle=\delta_{ii^\prime}$ . Since both $\{|i\rangle\}$ and $\{|e_i\rangle\}$ are orthonormal basis of the $d$ dimensional Hilbert space, and both $\{|i+(m-1)d\rangle\}$ and $\{|f_{im}\rangle\}$ are orthonormal basis of the $Kd$ dimensional Hilbert space, then there exist unitary operators $\tilde{U}_1$ and $\tilde{U_2}$ acting on the two subsystems respectively such that $\tilde{U}_1|f_{im}\rangle=|i+(m-1)d\rangle$ and $\tilde{U}_2|e_i\rangle=|i\rangle$ for all $i$ and $m$, which implies
\begin{eqnarray}
|\chi_m\rangle&=&\tilde{U}_1 \otimes \tilde{U}_2|\psi_m\rangle,\\
|\psi_m\rangle&=&\tilde{U}_1^\dagger \otimes \tilde{U}_2^\dagger|\chi_m\rangle,
\end{eqnarray}
for all $m$ and
\begin{eqnarray}
\tilde{U}_1 \otimes \tilde{U}_2 \rho \tilde{U}_1^\dagger \otimes \tilde{U}_2^\dagger =\sum_m p_m |\chi_m\rangle \langle \chi_m|.
\end{eqnarray}
Note that $I \otimes A |\chi_m\rangle= B^T \otimes I |\chi_m\rangle$, where $B$ is a block diagonal matrix
\begin{equation*}
B=\left(
\begin{array}{ccccc}
A & & & & \\
& A & & &\\
& & \ddots & &\\
& & & A & \\
& & & & I_{d^\prime-Kd}
\end{array}
\right).
\end{equation*}
Subsequently, $|\psi_m\rangle=\tilde{U}_1^\dagger G^T \otimes I |\chi_m\rangle$ with
\begin{equation*}
G=\left(
\begin{array}{ccccc}
\tilde{U}_2^\dagger & & & &\\
& \tilde{U}_2^\dagger & & &\\
& & \ddots & &\\
& & & \tilde{U}_2^\dagger &\\
& & & & I_{d^\prime-Kd}
\end{array}
\right)
\end{equation*}
for all $m$. Denote $U^\dagger= \tilde{U}_1^\dagger G^T$, then we have
\begin{equation}
|\psi_m\rangle=U^\dagger \otimes I|\chi_m\rangle
\end{equation}
for all $m$ and
$U\otimes I \rho U^\dagger\otimes I =\sum_{m=1}^K p_m |\chi_m\rangle \langle \chi_m|$.
Therefore, $\rho$ is maximally entangled if and only if there exists unitary
operator $U$ acting on the first subsystem such that Eq. (\ref{all max ent carnonical form }) holds.
\qed

Theorem \ref{th all max ent carnonical form} not only gives a necessary and sufficient condition for maximally entangled states, but also provides a canonical form for maximally entangled states. For example, in $2d\otimes d$ system, maximally entangled pure states can be transformed into
\begin{equation}
\frac{1}{\sqrt{d}}\sum_{i=0}^{d-1} |ii\rangle
\end{equation}
by unitary operations acting on the first subsystem. And maximally entangled mixed states can be transformed into
\begin{equation}
\frac{1}{d}(p_1 \sum_{i,j=0}^{d-1} |ii\rangle\langle jj| +p_2\sum_{i,j=0}^{d-1} |i+d,i\rangle\langle j+d,j|),
\end{equation}
$p_1, p_2>0$, $p_1+p_2=1$.

Now we give another necessary and sufficient condition for maximally entangled states.
\begin{theorem}\label{th mix max iff}
$\rho$ is maximally entangled if and only if there exists unitary operator $U$ acting on the first subsystem such that
\begin{eqnarray}\label{eq in th 1}
\sum_{m=1}^K \langle \chi_m| (U \otimes I)\,\rho \,(U^\dagger \otimes I) |\chi_m\rangle =1
\end{eqnarray}
with $|\chi_m\rangle$ defined in Eq. (\ref{max ent pure}).
\end{theorem}

Proof. For all maximally entangled state in Eq. (\ref{all max ent carnonical form }), it is easy to verify that $\sum_{m=1}^K \langle \chi_m|U\otimes I \rho U^\dagger \otimes I |\chi_m\rangle =\sum_{m=1}^K p_m=1$.

On the other hand, if a quantum state $\rho$ satisfies Eq. (\ref{eq in th 1}), then we only need to prove that
$\rho^\prime=U \otimes I\rho U^\dagger \otimes I$ is maximally entangled. Let $\rho^\prime=\sum_k \lambda_k |\phi_k\rangle \langle \phi_k|$, $\sum_k \lambda_k=1$, $\lambda_k>0$, be the spectral decomposition, then
\begin{eqnarray*}
&&\sum_{m=1}^K \langle \chi_m|U \otimes I\rho U^\dagger \otimes I |\chi_m\rangle \\
&=&\sum_{m=1}^K \langle \chi_m| \rho^\prime |\chi_m\rangle \\
&=&\sum_{m=1}^K \sum_k \lambda_k |\langle \chi_m |\phi_k\rangle |^2 \\
&=&1.
\end{eqnarray*}
Since $\sum_k \lambda_k=1$ and $\lambda_k>0$, one gets
\begin{eqnarray}\label{eq condi any pure with chi}
\sum_{m=1}^K |\langle \chi_m |\phi_k\rangle |^2=1
\end{eqnarray}
for all $k$.
Since $\{|\chi_m\rangle\}$ are orthonormal, we can extend them into a basis of $d^\prime d$ dimensional Hilbert space,
$\{|\chi_m\rangle,\ |\chi^\prime_t\rangle\}_{m,t}$. Under this basis, $|\phi_k\rangle$ can be expressed as $|\phi_k\rangle=\sum_m a_{mk} |\chi_m\rangle + \sum_t b_{tk} |\chi^\prime_t\rangle$ with $\sum_m |a_{mk}|^2+ \sum_t |b_{tk}|^2=1$ for all $k$. Taking into account
Eq. (\ref{eq condi any pure with chi}), which implies $\sum_m |a_{mk}|^2=1$, we get
\begin{eqnarray}\label{eq any pure}
|\phi_k\rangle&=&\sum_{m=1}^K a_{mk} |\chi_m\rangle\\\nonumber
&=&\frac{1}{\sqrt{d}}\sum_{i=0}^{d-1} |\xi_{ik}\rangle\otimes |i\rangle
\end{eqnarray}
with $|\xi_{ik}\rangle\equiv\sum_m a_{mk} |i+(m-1)d\rangle$ for all $k$.
From the orthonormality of the eigenvectors $|\phi_k\rangle$, we have
\begin{eqnarray}
\langle \phi_{k^\prime}|\phi_k\rangle&=&\sum_{m,m^\prime } a_{m^\prime k^\prime}^* a_{mk}\langle \chi_{m^\prime} |\chi_m\rangle\\\nonumber
&=&\sum_{m} a_{m k^\prime}^* a_{mk}\\\nonumber
&=&\delta_{kk^\prime}.
\end{eqnarray}
Hence
\begin{equation}
\begin{array}{rcl}\label{eq any pure sec sys}
&&\langle \xi_{i^\prime k^\prime}|\xi_{i k}\rangle\\&=&\sum_{m,m^\prime} a_{m^\prime k^\prime}^* a_{m k} \langle i^\prime +(m^\prime -1)d|i+(m-1)d\rangle\\[2mm]
&=& \sum_{m} a_{mk^\prime}^* a_{m k} \delta_{ii^\prime}\\[2mm]
&=&\delta_{kk^\prime}\delta_{ii^\prime}.
\end{array}
\end{equation}
Combining Eq. (\ref{eq any pure}) and Eq. (\ref{eq any pure sec sys}), one gets easily that the
quantum state $\rho^\prime $ is maximally entangled by Lemma \ref{lemma max}. Therefore $\rho$ is maximally entangled if and only if there exists unitary operator $U$ acting on the first subsystem such that
Eq. (\ref{eq in th 1}) holds.
\qed

\section{Fully entangled fraction}

The fully entangled fraction for
any quantum state $\rho$ in $d\otimes d$ system is defined as the maximal overlap with maximally entangled pure states,
\begin{equation}
\begin{array}{rcl}
F(\rho)&=&\max_{U,\ V} \langle \chi_1 |U\otimes V \rho U^\dagger \otimes V^\dagger |\chi_1 \rangle\\[2mm]
&=& \max_{U} \langle \chi_1|U\otimes I \rho U^\dagger\otimes I |\chi_1 \rangle.
\end{array}
\end{equation}
It measures how close a state is to maximally entangled states. The fully entangled fraction can be used to characterize whether a quantum state in $d\otimes d$ system can be used to teleport a $d$ dimensional quantum state faithfully. But one shortage of this quantity is that it is only well defined in $d\otimes d$ system and it does not make sense in quantum systems with different dimensional subsystems.

To deal with this matter, we propose the fully entangled fraction of quantum state $\rho$ in $d^\prime \otimes d$ system as
\begin{equation}\label{def gen fully frac}
\begin{array}{rcl}
F(\rho)&=& \max_{U,V}\displaystyle \sum_{m=1}^K \langle \chi_m |(U \otimes V)\, \rho\,( U^\dagger \otimes V^\dagger) |\chi_m\rangle\\[4mm]
&=&\max_U \displaystyle\sum_{m=1}^K \langle \chi_m |(U \otimes I)\, \rho \,(U^\dagger \otimes I) |\chi_m\rangle,
\end{array}
\end{equation}
where $d^\prime=Kd+r$, $0\leq r<d$, $K\geq 1$.

$F(\rho)$ has the following properties.

(1) $F(\rho)$ is invariant under local unitary transformations.

(2) $F(\rho)$ is linear and convex.

(3) $F(\rho)=1$ if and only if $\rho$ is maximally entangled.

(4) $\frac{K}{d^\prime d}\leq F(\rho)\leq 1$ for all $d^\prime \otimes d$ quantum states $\rho$. Especially for $d \otimes d$ mixed state $\rho$, $\frac{1}{d^2}\leq F(\rho)\leq 1$.

(5) $\frac{K}{d^\prime d}\leq F(\rho)\leq \frac{1}{d}$ for all $d^\prime \otimes d$ separable states $\rho$.

Since the first three properties are easy to derive, here we only prove the last two properties.

Proof of property (4).
For any $d^\prime \otimes d$ mixed state $\rho$, we assume
$\rho=\sum_{i=1}^{d^\prime d} \lambda_i |\phi_i\rangle \langle \phi_i|$ be
the spectral decomposition such that $\sum_{i=1}^{d^\prime d}\lambda_i=1$,
$0 \leq \lambda_i \leq 1$ and $\{ |\phi_i\rangle \}_{i=1}^{d^\prime d}$ are
the normalized orthogonal eigenvectors in $d^\prime \otimes d$ Hilbert space.
Then ${F}(\rho)=\max_U \sum_{i=1}^{d^\prime d} \lambda_i a_i$,
with $a_i = \sum_{m=1}^K \langle \chi_m |U \otimes I |\phi_i\rangle \langle \phi_i| U^\dagger \otimes I |\chi_m\rangle$, which satisfies $0 \leq a_i \leq
1$ and $\sum_{i=1}^{d^\prime d} a_i=K$. One gets that $\sum_{i=1}^{d^\prime d} \lambda_i a_i
\leq \sum_{i=1}^{d^\prime d} \lambda_i =1$ becomes an equality if and only
if $a_i=1$ for all $i$. Therefore $F(\rho)=1$
if and only if $\rho$ is maximally entangled state.

On the other hand, the minimum of the function $g(\lambda_i,
a_i)=\sum_{i=1}^{d^\prime d} \lambda_i a_i$ is $\frac{K}{d^\prime d}$ by Lagrange
multiplier. It reaches its minimum if and only if $\lambda_i=\frac{1}{d^\prime d}$ and $
a_i=\frac{K}{d^\prime d}$ for $i=1, \cdots, d^\prime d$. This gives rise to
$\rho=\frac{1}{d^\prime d}I$. \qed

Proof of property (5). For $d^\prime \otimes d$ pure separable state $|00\rangle$, $F(|00\rangle)=\max_U \sum_{m=1}^K \langle \chi_m |U \otimes I |00\rangle \langle 00| U^\dagger \otimes I |\chi_m\rangle=\frac{1}{d}\max_U \sum_{m=1}^K  |U_{(m-1)d,0}|^2\leq \frac{1}{d}$. Notice that fully entangled fraction is local unitary invariant and convex, we know $F(\rho)\leq \frac{1}{d}$ for all $d^\prime \otimes d$ separable states. \qed

One plausible weakness of the fully entangled fraction is that for any given quantum state, its fully entangled fraction may depend on the dimension of the Hilbert space associated to the state. For example, $F_1(|00\rangle\langle00|)=\frac{1}{d}$ for $|00\rangle\langle00|$ in $d\otimes d$ Hilbert space. Therefore, the fully entangled fraction of $|00\rangle\langle00|$ is $\frac{1}{2}$ if we consider it as a $2\otimes 2$ state and $\frac{1}{3}$ if we consider it as a $3\otimes 3$ state. So before calculating the fully entangled fraction, we first need to identify the associated Hilbert space.
This is also the problem that the usual fully entangled fraction $F(\rho)$ should be confront with. However, the problem is not so serious. The reason is that
once we say how much entanglement one quantum state has, we are subconsciously comparing this state with others. If one state is maximally entangled, $F(\rho)=1$, then it means that it has more entanglement than others with $F(\rho)<1$ in the same Hilbert space. So for $\rho_1$ and $\rho_2$ in the same Hilbert space, if $F(\rho_1)>F(\rho_2)$, it implies that $\rho_1$ has more entanglement than $\rho_2$ and $\rho_1$ may be more useful in some sense.

Now we show the roles played by $F(\rho)$ in the following quantum teleportation. Suppose Alice and Bob previously share a pair of particles in arbitrary $d^\prime \otimes d$ quantum state $\rho$. To transform an unknown $d$ dimensional state $|\psi\rangle$ from Alice to Bob, Alice first performs a generalized joint Bell measurement $|\phi_{s,t,m}\rangle \langle \phi_{s,t,m}|$ with $|\phi_{s,t,m}\rangle=U_{st}\otimes I (\frac{1}{\sqrt{d}}\sum_i |i\rangle|i+(m-1)d\rangle)$ on her parties, here $U_{st}=h^t g^s$, $h|j\rangle=|(j+1)\mod d\rangle$, $g|j\rangle=\omega^j |j\rangle$ with $\omega=exp\{-2i\pi/d\}$, $s,~t=1,2,\cdots,d$, $m=1,\cdots, K$. According to the measurement results $s,~t,~m$ of Alice, Bob chooses particular unitary transformations $T_{s,t,m}$ to act on his particle. By lengthy calculation, we find the transmission fidelity of the teleportation protocol defined by $T_{s,t,m}$ is given by
\begin{eqnarray*}
f(\rho)=&&\frac{1}{d+1}\\&&+ \frac{1}{d(d+1)}\sum_{s,t,m} \langle \chi_m| (I \otimes U_{st}^\dagger T_{stm}^\dagger)\, \rho\, (I\otimes T_{stm} U_{st}) |\chi_m\rangle.
\end{eqnarray*}
So the optimal fidelity is
\begin{equation}
\frac{1}{d+1}(1+\max_V \sum_{m} \langle \chi_m| (I \otimes V)\, \rho \, (I\otimes V^\dagger) |\chi_m\rangle),
\end{equation}
where $V$ are arbitrary $d\times d$ unitary operators.

If we allow local unitary operations first before the above quantum teleportation, then
it gives raise to the relation between the optimal fidelity of the teleportation and the fully entangled fraction,
\begin{eqnarray}
f_{\max}(\rho)=\frac{1}{d+1}+ \frac{d F(\rho)}{d+1}.
\end{eqnarray}

For separable states $\rho$, $F(\rho)\leq \frac{1}{d}$, the optimal fidelity $f_{\max}(\rho)$ of $\rho$ in quantum teleportation is no more than $\frac{2}{d+1}$. However if $F(\rho)>\frac{1}{d}$, then its optimal fidelity is not less than $\frac{2}{d+1}$ and $\rho$ is useful for quantum teleportation. In this sense, the fully entangled fraction $F(\rho)$ can be used to detect quantum teleportation resource.

\section{Entanglement witness}

To detect maximally entangled states and quantum teleportation resource in general bipartite system experimentally, we construct the following entanglement witness. First, let us define the Hermitian operators,
\begin{widetext}
\begin{eqnarray*}
\lambda_{i+(m-1)d}&=&|(m-1)d\rangle\langle (m-1)d| - |i+(m-1)d\rangle\langle i+(m-1)d|,\\[2mm]
\lambda_{k+(m-1)d,\ l+(m-1)d}&=&|k+(m-1)d\rangle\langle l+(m-1)d| + |l+(m-1)d\rangle\langle k+(m-1)d|,\\[2mm]
\lambda_{k+(m-1)d,\ l+(m-1)d}^\prime&=&{\rm i}(|k+(m-1)d\rangle\langle l+(m-1)d| - |l+(m-1)d\rangle\langle k+(m-1)d|),
\end{eqnarray*}
for the first subsystem,
and
\begin{eqnarray*}
\lambda_i&=&|0\rangle\langle0| - |i\rangle\langle i|,\\[2mm]
\lambda_{kl}&=&|k\rangle\langle l| + |l\rangle\langle k|,\\[2mm]
\lambda_{kl}^\prime&=&{\rm i}(|k\rangle\langle l| - |l\rangle\langle k|),
\end{eqnarray*}
for the second subsystem, with $i=1,\cdots,d-1$; $k,l=0,\cdots,d-1$, $k<l$; $m=1,\cdots, K$.

Furthermore, let
\begin{eqnarray*}
\mu_{i}&=&\sum_{m=1}^K \lambda_{i+(m-1)d},\\[2mm]
\mu_{kl}&=&\sum_{m=1}^K \lambda_{k+(m-1)d,\ l+(m-1)d},\\[2mm]
\mu_{kl}^\prime&=&{\rm i}\sum_{m=1}^K \lambda_{k+(m-1)d,\ l+(m-1)d}^\prime.
\end{eqnarray*}
Set $A_{i}= U\mu_{i} U^\dagger$, $A_{kl}=U \mu_{kl} U^\dagger$, $A_{kl}^\prime=U \mu_{kl}^\prime U^\dagger$, with $U$ any $d^\prime\times d^\prime$ unitary matrix.
We define the linear witness operator to be
\begin{equation}\label{eq mix max operator}
\begin{array}{rcl}
\displaystyle\Gamma\equiv \displaystyle\frac{1}{d^2}[ I_{d^\prime} \otimes I_d
+d\sum_{i=1}^{d-1} A_{i} \otimes \lambda_i - \sum_{i=1}^{d-1}\sum_{j=1}^{d-1}  A_{i}\otimes \lambda_j]
\displaystyle +\frac{1}{2d}\sum_{0\leq k< l\leq d-1}[  A_{kl} \otimes \lambda_{kl}  - A_{kl}^\prime \otimes \lambda_{kl}^\prime ].
\end{array}
\end{equation}

\begin{theorem}
$\rho$ is maximally entangled if and only if
\begin{eqnarray}\label{eq witness max ent}
\langle \Gamma\rangle_{\rho}=1,
\end{eqnarray}
and it is useful in quantum teleportation if and only if
\begin{eqnarray}\label{eq witness tele}
\langle \Gamma\rangle_{\rho}>\frac{1}{d}
\end{eqnarray}
for some unitary operator $U$ acting on the first subsystem.
\end{theorem}

Proof. By expanding the operator $|\chi_m\rangle\langle\chi_m|$ in Eq. (\ref{max ent pure}) in terms of
the Hermitian operators $\lambda_{i+(m-1)d}$, $\lambda_{k+(m-1)d,\ l+(m-1)d}$, $\lambda_{k+(m-1)d,\ l+(m-1)d}^\prime$ on the first subsystem, and $\lambda_i$, $\lambda_{kl}$, $\lambda_{kl}^\prime$ on the second subsystem, $i=1,\cdots,d-1$; $k,l=0,\cdots,d-1$, $k<l$; $m=1,\cdots, K$, we have
\begin{eqnarray}\label{chi m operator form}
|\chi_m\rangle\langle\chi_m|&=&\displaystyle\frac{1}{d^2}[ I_{d_{m-1}} \otimes I_d
+d\sum_{i=1}^{d-1} \lambda_{i+(m-1)d}\otimes \lambda_i - \sum_{i=1}^{d-1}\sum_{j=1}^{d-1}  \lambda_{i+(m-1)d}\otimes \lambda_j]
\displaystyle \\\nonumber&&+\frac{1}{2d}\sum_{0\leq k< l\leq d-1}[ \lambda_{k+(m-1)d,\ l+(m-1)d} \otimes  \lambda_{kl} - \lambda_{k+(m-1)d,\ l+(m-1)d}^\prime \otimes \lambda_{kl}^\prime].
\end{eqnarray}
\end{widetext}
Inserting Eq. (\ref{chi m operator form}) into $\sum_{m=1}^K U\otimes I |\chi_m\rangle\langle\chi_m|U^\dagger\otimes I$, one gets $\rho$ is maximally entangled if and only if
Eq. (\ref{eq witness max ent}) holds
for some unitary operator $U$ acting on the first subsystem by theorem \ref{th mix max iff}, and $\rho$ is useful for quantum teleportation if and only if
Eq. (\ref{eq witness tele}) holds
for some unitary operator $U$ acting on the first subsystem, by the result in section III.
\qed

For example, in $4\otimes 2$ system, the witness given by Eq. (\ref{eq mix max operator}) is
\begin{equation}
\begin{array}{rcl}
\Gamma=&&\frac{1}{4}[I_4\otimes I_2 + U\lambda_1U^\dagger\otimes \lambda_1 + U\lambda_3U^\dagger\otimes\lambda_1 \\[2mm]
&& + U\lambda_{01}U^\dagger\otimes\lambda_{01}+ U\lambda_{23}U^\dagger\otimes\lambda_{01}\\[2mm]
&&-U\lambda_{01}^\prime U^\dagger\otimes\lambda_{01}^\prime - U\lambda_{23}^\prime U^\dagger\otimes\lambda_{01}^\prime].
\end{array}
\end{equation}
Then any quantum state is maximally entangled if and only if $\langle \Gamma\rangle_{\rho}=1$, and the state
is useful for quantum teleportation if and only if $\langle \Gamma\rangle_{\rho}>\frac{1}{2}$.

\section{Conclusions}

In summary, we have studied maximally entangled states and fully entangled fraction in general $d^\prime\otimes d$ systems.
We have presented necessary and sufficient conditions of the maximally entangled states.
The maximal overlap between a given quantum state and the maximally entangled states has been characterized
by the fully entangled fraction, analogous to the case in $d\otimes d$ systems,
as a natural generalization of usual fully entangled fraction. The properties of the fully entangled fraction
and its relation to quantum teleportation have been analyzed. These investigation completes the previous results for fully entangled fraction.
The witness for detecting the maximally entangled states and the resource for quantum teleportation has provided,
which may be helpful for experimental detection.

\section{Acknowledgement}

This work is supported by NSF of China under Grant No. 11401032 and No. 61473325.

\end{document}